\documentclass[useAMS,usenatbib]{mn2e}  
\usepackage{amsmath,amsfonts,epsfig,natbib,mathptm}

\def\reff@jnl#1{{\rm#1\/}}

\def\aj{\reff@jnl{AJ}}                  
\def\araa{\reff@jnl{ARA\&A}}            
\def\apj{\reff@jnl{ApJ}}                        
\def\apjl{\reff@jnl{ApJ}}               
\def\apjs{\reff@jnl{ApJS}}              
\def\ao{\reff@jnl{Appl.Optics}}         
\def\apss{\reff@jnl{Ap\&SS}}            
\def\aap{\reff@jnl{A\&A}}               
\def\aapr{\reff@jnl{A\&A~Rev.}}         
\def\aaps{\reff@jnl{A\&AS}}             
\def\azh{\reff@jnl{AZh}}                        
\def\baas{\reff@jnl{BAAS}}              
\def\jrasc{\reff@jnl{JRASC}}            
\def\memras{\reff@jnl{MmRAS}}           
\def\mnras{\reff@jnl{MNRAS}}            
\def\pra{\reff@jnl{Phys.Rev.A}}         
\def\prb{\reff@jnl{Phys.Rev.B}}         
\def\prc{\reff@jnl{Phys.Rev.C}}         
\def\prd{\reff@jnl{Phys.Rev.D}}         
\def\prl{\reff@jnl{Phys.Rev.Lett}}      
\def\pasp{\reff@jnl{PASP}}              
\def\pasj{\reff@jnl{PASJ}}              
\def\qjras{\reff@jnl{QJRAS}}            
\def\skytel{\reff@jnl{S\&T}}            
\def\solphys{\reff@jnl{Solar~Phys.}}    
\def\sovast{\reff@jnl{Soviet~Ast.}}     
\def\ssr{\reff@jnl{Space~Sci.Rev.}}     
\def\zap{\reff@jnl{ZAp}}                        
\def\nat{\reff@jnl{Nature}}             

\title[The CMB power spectrum measured by the VSA.]{The CMB power spectrum out
to $\ell = 1400$ measured by the VSA}

\author[Keith Grainge et al.]  {Keith Grainge$^1$, Pedro Carreira$^2$,
  Kieran Cleary$^2$, Rod D. Davies$^2$, Richard J.  Davis$^2$,
  \newauthor Clive Dickinson$^2$, Ricardo Genova-Santos$^3$, Carlos M.
  Guti{\'e}rrez$^3$, Yaser A.
  Hafez$^2$, \newauthor Michael P. Hobson$^1$, Michael E.
  Jones$^1$, R\"udiger Kneissl$^1$, Katy Lancaster$^1$,\newauthor
 Anthony Lasenby$^1$, J. P. Leahy$^2$, Klaus
  Maisinger$^1$, Guy G. Pooley$^1$, 
  Rafael Rebolo$^{3,4}$,  \newauthor Jos\'e Alberto
  Rubi\~no-Martin$^{3,\ddagger}$,  Pedro Sosa Molina$^3$, Carolina
  \"{O}dman$^1$, 
  Ben Rusholme$^{1,\star}$,\newauthor  Richard D.E. Saunders$^1$, 
  Richard Savage$^1$, Paul F. 
  Scott$^1$, An\v ze Slosar$^1$,  Angela C. Taylor$^1$, \newauthor
  David Titterington$^1$, Elizabeth Waldram$^1$, Robert A.
  Watson$^{2,\dagger}$,  Althea Wilkinson$^2$
  \\
  $^1$ Astrophysics Group, Cavendish Laboratory, University of Cambridge, UK\\
  $^2$ Jodrell Bank Observatory, Macclesfield, Cheshire, SK11 9DL, UK\\
  $^3$ Instituto de Astrof{\'i}sica de Canarias, 38200 La Laguna,
  Tenerife, Spain.\\
  $^4$Consejo Superior de Investigaciones Cient{\'{\i}}ficas, Spain \\
  $^{\star}$Present address: Stanford University, Palo Alto, CA,
  USA\\
  $^{\dagger}$Present address: Instituto de Astrof{\'{\i}}sica de
Canarias.\\
  $^{\ddagger}$Present address: Max-Planck Institut f\"ur Astrophysik,
  Garching, Germany}

  \pagerange{\pageref{firstpage}--\pageref{lastpage}}

\pubyear{2002}

\begin{document}
\maketitle
\label{firstpage}
\begin{abstract}

We have observed the cosmic microwave background (CMB) in three regions of sky
using the Very Small Array (VSA) in an extended configuration with antennas
of beamwidth $2^{\circ}$ at 34~GHz. Combined with data from previous
VSA observations using a more compact array with larger beamwidth, we measure
the power spectrum of the primordial CMB anisotropies between angular 
multipoles $\ell = 160$ -- $1400$. 
Such
measurements at high $\ell$ are vital for breaking degeneracies in parameter
estimation from the CMB power spectrum and other cosmological data.
The power spectrum clearly resolves the first three acoustic
peaks, shows the expected fall off in power at high $\ell$
and starts to constrain the position and height of a fourth peak.

\end{abstract}


\section{Introduction}

Acoustic peaks in the power spectrum of cosmic microwave
background~(CMB) anisotropies were predicted by \citet{sakharov},
\citet{sunyaev-70} and \citet{peebles-70}. 
Recently several experiments have accurately measured the first of these peaks
and have detected the second peak
~\citep{lee-01,netterfield-01,halverson-02,VSApaperIII}.  Importantly, despite
the differing types of potential systematic errors suffered by these
experiments, they are in very good agreement with each other. The CBI
experiment~\citep{padin-02} has measured the power spectrum out to $\ell \
\sim \ 3500$~\citep{mason-02,pearson-02}, and has detected the predicted fall
in power level due to incoherent addition of temperature fluctuations
along the line of sight and Silk damping~\citep{silk}, but does not have the
$\ell$-resolution to define the peak structure. Resolving the third
and subsequent 
CMB peaks is essential to constrain further cosmological models.  The Very
Small Array (VSA)~\citep{VSApaperI} (hereafter Paper\,I) is an interferometric
array which has measured the CMB power spectrum between $ \ell \ = $ 150--800
as described in \citet{VSApaperII} and \citet{ VSApaperIII} (hereafter
Paper\,II and Paper\,III respectively).  These measurements were made with the
VSA's 14 antennas arranged in a compact configuration. The VSA has since been
upgraded with larger apertures to allow observations in an extended
configuration giving high temperature-sensitivity measurements on smaller
angular scales. This paper describes the first VSA measurements with the
extended array and the power spectrum from the combined compact and extended
arrays.

\begin{table}
\begin{center}
 \caption{VSA field positions and total effective integration time
   remaining after flagging and filtering of the
   data.\label{tab:fieldpos}}
  \begin{tabular}{lccc}
    \hline
     & RA (J2000) & DEC (J2000)& $\rm T_{int}$ (hrs)\\
       \hline
VSA1E     &     00 22 37  & 30 16 38 & 106\\
VSA1F     &     00 16 52  & 30 24 10 & 94\\
VSA1G     &     00 19 22  & 29 16 39 & 79\\
VSA2E     &     09 37 57  & 30 41 28 & 110\\
VSA2F     &     09 43 46  & 30 41 14 & 101\\
VSA2G     &     09 40 53  & 31 46 21 & 115\\
VSA3E     &     15 31 43  & 43 49 53 & 130\\
VSA3F     &     15 38 38  & 43 50 18 & 114\\
VSA3G     &     15 35 13  & 42 45 05 & 112\\
\hline
\end{tabular}
\end{center}
\end{table}

\section{The VSA extended array}

As described in Paper\,I, each VSA antenna comprises a conical corrugated horn
feeding a section of a paraboloidal mirror.  In its extended configuration,
the 143 mm diameter illuminated apertures of the compact array are replaced
with 322 mm diameter apertures, giving a primary beam of $2.0^{\circ}$ FWHM at
34~GHz and an improvement in flux sensitivity of a factor of just over 5. The
filling factor of this configuration is greater than that of the
compact array, giving 
a significant increase in the overall temperature sensitivity.

The extended array has several features which increase observational
efficiency. The narrower primary beam response allows observations closer to
the Sun and the Moon than for the compact array. 
Repeating the tests outlined
in Paper\,I we conclude observations must be at least
$18^{\circ}$ from the Sun and the Moon.
Also the higher gain and directivity of the new
antennas results in a reduced level of cross-coupling, now measured to be less
than $-120$~dB. The `spurious signal' discussed in Paper\,I has now largely
vanished, even on the shortest baselines, but conservatively we
still apply a high-pass fringe-rate filter to our data to remove any
low-level contamination.

Early commissioning runs with the extended array showed that, at the extremes
of the VSA's pointing range, the data were contaminated when sidelobes of the
telescope's primary beam were directed towards the top edge of the VSA ground
screen; this edge appears as a temperature discontinuity and so results in a
correlated signal. Tests show that the contamination is negligible for hour
angles of less than $\pm 3$~hrs; adopting a conservative approach
we limit observations 
with the VSA extended array to hour angles of $\pm 2.5$~hrs.

\section{Observations}

\subsection{Field centres}

The observations presented in this paper were made at 34~GHz during the period
2001 October -- 2002 April.  In each of three regions of sky, three
overlapping fields were observed, separated by 75 arcminutes. All the fields
lie within the FWHM of the mosaiced fields already observed at lower
resolution by the VSA in its compact configuration~\citep{VSApaperII}. The
positions and effective integration times of the nine fields are given in
Table \ref{tab:fieldpos}.  The fields were selected for low synchrotron and
free-free emission, low dust and a lack of large-scale structure and are known
from previous observations to contain no radio sources brighter
than 100~mJy at 34~GHz. In Paper\,II we used external foreground
templates for synchrotron, free-free and the dust, to estimate the level of
contamination in the VSA fields.  We showed that the amplitude of the combined
foreground signal at $1^{\circ}$ angular scales is $< 6~\mu$K.  It is known
that the power spectra of all the Galactic foregrounds decreases with
increasing $\ell$ with power-law index $\alpha = $ 2--3 ($C_{\ell} \propto
\ell^{-\alpha}$)~\citep{giardino01,tegmark00}.
Assuming a power-law index of 2.5, at $\ell=1200$ the Galactic
foregrounds will contribute no more than $\approx 16~\mu\rm{K^2}$ to
the power spectrum.  We therefore conclude that contamination from
Galactic emission will be negligible compared to our random
errors. Foreground contamination is discussed further by Dickinson et
al. (in prep).

\subsection{Data reduction}

The data were reduced and analysed as described in detail in Papers I
and II but with three differences.

\begin{description}

\item{\bf Calibration.} Our primary calibrator was Jupiter, whose effective
temperature at 34~GHz is $T_{34}=155\pm5$~K~\citep{mason_casscal}. We also use
Cas~A and Tau~A as phase calibrators, transferring the flux scale from
Jupiter. These sources are partially resolved on the longer baselines
of the VSA 
extended array and it is necessary to model them. We used VLA 1.4~GHz images
\citep{CasA,Bietenholz91} with aperture plane coverage encompassing
that of the VSA 
to model the change in observed flux density over the range of VSA
baselines. Both Cas~A and Tau~A are sufficiently similar in structure at 1.4
and 34~GHz for this process to work well; we have confirmed this using a
32-GHz image from Effelsberg for Cas~A (W. Reich priv comm) and an 850-$\mu$m
SCUBA map for Tau~A \citep{Crab}. We find that the amplitude of the
correction is about 15 percent on the longest VSA baselines.

\item{\bf Effects of bad weather.} Some of the extended array data were taken
in weather conditions considerably worse than those for the compact
array. Days when the sky temperature increased significantly were identified
from the VSA system temperature monitor and the data were discarded. However,
isolated clouds unresolved by the telescope beam can give correlated signals
in the visibilities without any detectable increase in the system
temperature. This emission is visible in the data as periods of very rapidly
varying fringes. These fluctuations are generally short-lived and strongest on
projected baselines which are perpendicular to the wind direction. To detect
this correlated emission we calculate the variance of the 1-second sampled
data over all 91 baselines during a 16-second integration; this is then used
to reweight the data for each sample. This is in addition to weighting for
system temperature variations and antenna effective sensitivities.

\item{\bf Correction for fringe smearing.} To remove DC correlator offsets we
apply a phase-switching sequence in the hardware with a cycle time of 16~s,
which is then demodulated in software. For the higher fringe rates of the
extended array, this process results in a reduction in the signal amplitude by
a factor of up to $6\%$ and we apply an appropriate correction and
downweighting of the data for each visibility.

\end{description}

The data were analysed independently by each of the three collaborating
institutions, and the results were found to be consistent to within a small
fraction of the intrinsic uncertainties in the data. In contrast to the
compact array analysis, only 28 percent of the data were discarded due to
weather, fringe rate filtering and telescope downtime; this difference is due
in part to the higher fringe rates of the extended array.

\subsection{Source subtraction} 

The point source contribution to the observed power spectrum
increases with $\ell$ as
\mbox{$T_0^2\,C_0\ell(\ell+1)/2\pi$}, and contamination from
extragalactic radio sources is expected to be significant for
observations made using the VSA extended array.  Our source removal
strategy is to survey the fields at 15~GHz with the Ryle
Telescope~(RT) and then follow up the sources detected using a
single-baseline source-subtraction interferometer operating at
34~GHz. The source-subtractor baseline is just longer than
$1000\lambda$ and so the contamination from primordial fluctuations is
negligible. 

To ensure that extended-array observations of the CMB are not compromised by
point-source contamination, it is necessary to subtract fainter sources than
was required for the compact array. Taking the 34~GHz source count derived in
Paper\,II, if we subtracted sources to the same level as before (i.e. S$_{34}
> 80$~mJy), then the residual source power at $\ell=1500$ would be ${\Delta
T}^2_{\rm sources} \sim 1800 \, \mu{\rm K}^{2}$. This is significantly higher
than the predicted CMB power spectrum. For the current observations we
subtract sources down to S$_{34} > 20$~mJy, which limits the contribution of
unsubtracted sources to less than $\sim 470\,\mu{\rm{K}^{2}}$ at
$\ell~=~1500$,  
and provides
better statistics for the removal of this residual
contribution to the power spectrum.

To achieve this levl of source removal, several changes were made to
the source subtraction strategy. 
The flux limit of the RT survey at 15~GHz was
improved to 10~mJy, which ensures that sources with a spectral index as steep
as $-1$ between 15 and 30~GHz can be found \citep{waldram-02}. The
sensitivity of the source subtractor was also improved by replacing the dishes
with ones of increased surface accuracy and hence higher aperture efficiency,
and by upgrading to a double sideband system, doubling the effective
bandwidth.

In total 24 sources were detected above 20~mJy at 34~GHz and removed.
We found reasonable upper and lower limits to the source count in the
range 20 -- 100 mJy to be $N(S) dS = 7.6 \times 10^3 (S/100 \rm
mJy)^{-2.15} \rm \,Jy^{-1}\,sr^{-1}$ and $N(S) dS = 8.5 \times 10^3
(S/100 \rm mJy)^{-1.8}\,\rm Jy^{-1}\,sr^{-1}$ (Taylor et al. in
prep. will present a full discussion of the 34~GHz
counts). Integrating these counts from zero to 20~mJy gives a range of
residual source powers of $T_0^2\,C_0=1.3\pm 0.5 \times 10^{-3} \, \mu
\rm K^2$. We therefore subtract this statistical correction from the
power spectrum band powers, in addition to the subtraction of discrete
sources from the visibilities.

\section{Results}

\begin{figure*}
\begin{minipage}{150mm}
\epsfig{file=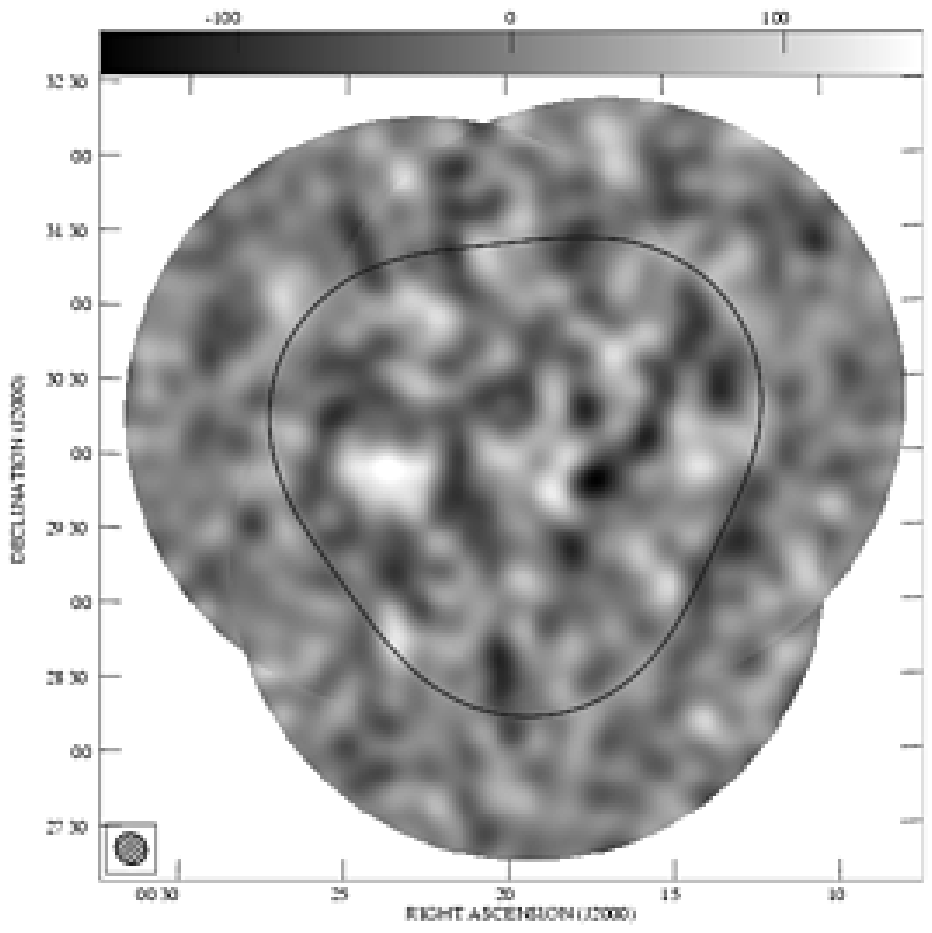,angle= 0,width=5cm}
\epsfig{file=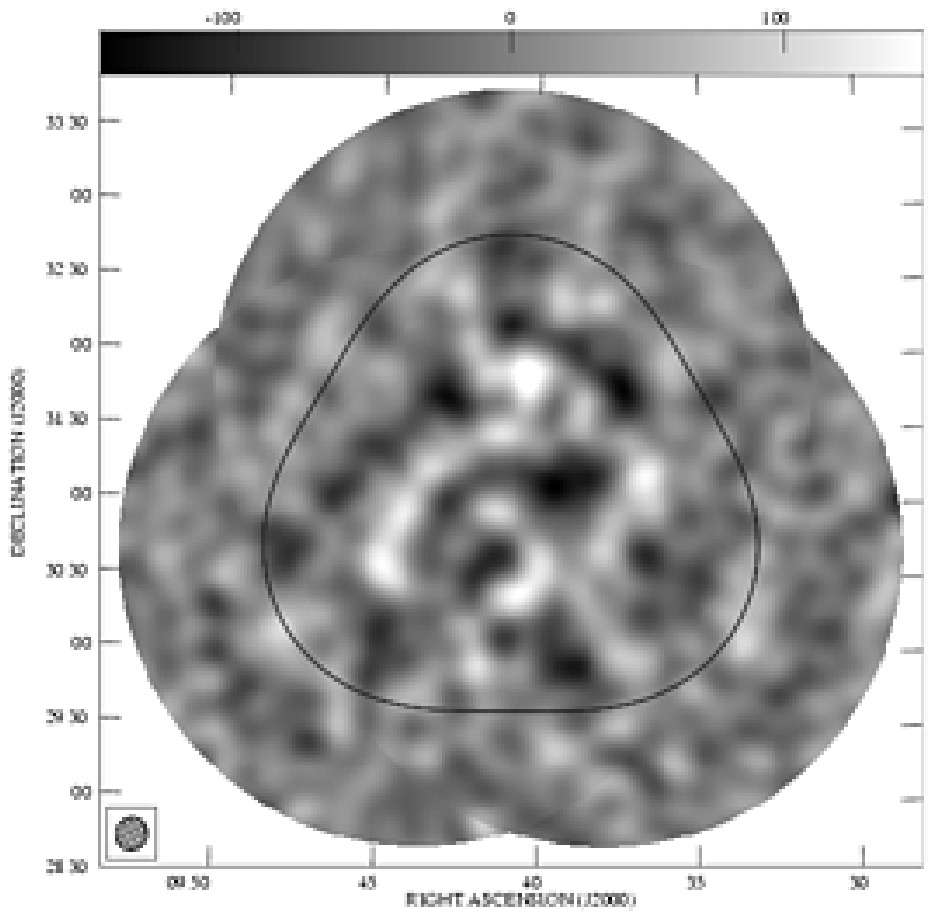,angle= 0,width=5cm}
\epsfig{file=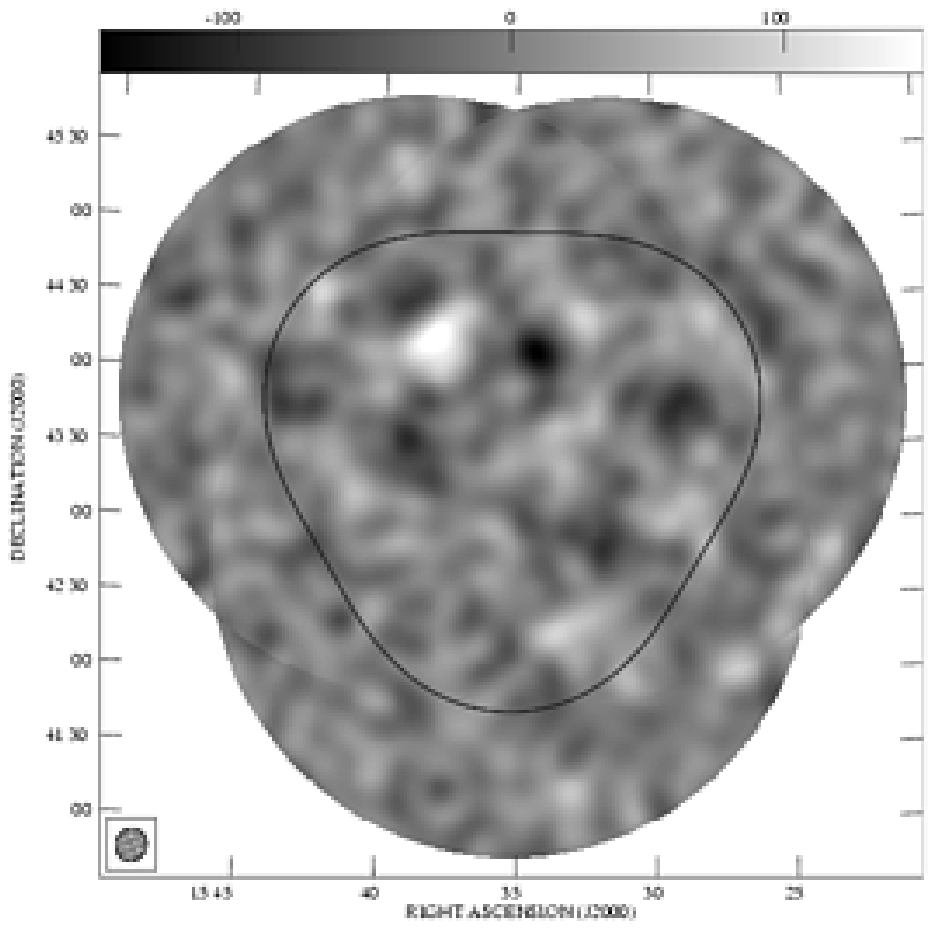,angle= 0,width=5cm}
\caption{Mosaiced and source-subtracted maps for each of the VSA regions:
left, VSA1; centre, VSA2; right, VSA3. In each case the greyscale runs from
$-150$ to $150 \, \mu$K and the contour indicates the half-power response of the
combined primary beam.
\label{fig:maps}}
\end{minipage}
\end{figure*}

\begin{figure}
\epsfig{file=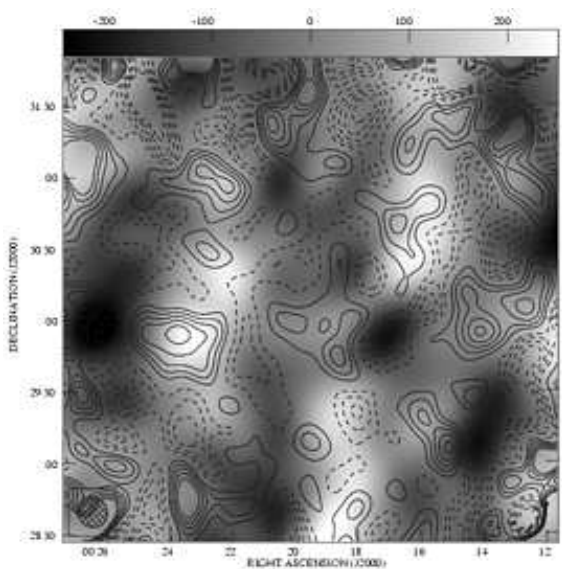,angle=0,width=7.5cm}
\caption{Comparison of the compact and extended array maps for the
central part of the 
VSA1 field. Greyscale: Compact array map, with greyscale running from $-250$
to $250 \, \mu$K. The sensitivity is nearly uniform across this
area. Contours: Extended array map, contour interval $30 \, \mu$K. The map has
been corrected for the combined effect of the primary beams, so the noise
level rises towards the edge of the map. 
\label{fig:comp-map}}
\end{figure}

In Figure \ref{fig:maps} we present maps of the three regions observed. These
maps provide important data checks and allow comparison with the earlier maps
of the same sky areas made using the compact array but are not used in the
calculation of the power spectrum, which is computed directly from the
visibility data. Individual maps were made of each of the nine pointings,
using a radial weighting function in the aperture plane proportional to
$C_{\ell}^{1/2}$ of a fitted $\Lambda$CDM power spectrum
-- this is equivalent to a Wiener filter. The maps were
CLEANed to reduce the long-distance correlations arising from the restricted
sampling in the aperture plane and then combined into mosaics, weighted by
their respective primary beams, using the AIPS task
{\sc{Ltess}}. Figure~\ref{fig:comp-map} shows a comparison between the
compact and extended array 
maps of the VSA1 field; the two maps have different but overlapping aperture
plane coverages and show many features in common.

We find that the noise levels in our maps well
outside the primary beam are consistent with the thermal noise level derived
from subtracting alternate visibilities ($\sigma_{\rm auto}$), showing that
there is no component in the maps other than the signal within the primary
beam and the thermal noise.
The thermal noises for each field are given in  
Table~\ref{tab:noises} together with the rms CMB contributions in
the central region of each of the maps.

\subsection{Data checks}

\begin{table}
\begin{center}
 \caption{Rms noise levels, in mJy beam$^{-1}$, for each of the VSA fields: 
the thermal noise level, $\sigma_{\rm auto}$ (measured from the
autosubtracted maps -- see Paper\,II); the noise level in the centre of
   each map, $\rm \sigma_{C}$; the rms CMB signal, $\rm
   \sigma_{CMB}$.
   These values are taken from CMB maps made at full resolution. The
   flux-to-temperature conversion is about $\rm 1 \, mJy \, beam^{-1}
   \equiv 2.1 \, \mu K$, giving mean rms CMB fluctuations in the maps
   of $\approx 
   30 \,\mu$K.\label{tab:noises} } \begin{tabular}{lccc} \hline
Field  & $\sigma_{\rm auto}$ & $\rm \sigma_{C}$ &
       $\sigma_{\rm CMB}$ \\
       \hline
  VSA1E & 11 & 19 & 15\\
  VSA1F & 11 & 19 & 15\\
  VSA1G & 11 & 15 & 10\\
  VSA2E & 10 & 19 & 15\\
  VSA2F & 11 & 22 & 18\\ 
  VSA2G & 10 & 23 & 19\\
  VSA3E & 10 & 18 & 14\\
  VSA3F & 10 & 20 & 15\\  
  VSA3G & 9  & 14 & 10\\ 
  \hline
 \end{tabular}
\end{center}
\end{table}

As a check on the consistency of the data, we computed the $\chi^{2}$
statistic and significance 
for splits on the visibility data following the method
described in Paper\,III. The significance is
given as the probability to exceed the observed value in the
$\chi^2$ cumulative distribution function.
The data for each VSA field were split in two according to observing
epoch and the $\chi^2$ values and associated significances are given
in Table~\ref{chi}. A split between day and night observations is also
included.

The consistency of the power spectra derived from each of the 3 VSA
mosaiced regions was compared by forming the $\chi^2$ statistic
on pairs of power spectra.
The $\chi^{2}$ values (and significances) for the VSA1/VSA2, VSA1/VSA3 and
VSA2/VSA3 power spectra comparisons are 12.2 (0.73), 9.5 (0.89) and 15.4
(0.50) respectively. In each case there are 16 degrees of freedom in the power
spectrum analysis. The data 
sets are clearly self-consistent.

\begin{table}
\caption{The $\chi^2$ values for data splits on each of the
  VSA fields.  In each case the visibility data from each field were
  split in two according to epoch and the $\chi^2$ of the difference
  vectors formed.  Also tabulated are the number of degrees of freedom
  (DOF) and the significance of each $\chi^2$ value. 
\label{chi}}
\begin{center}
\begin{tabular}{|c|c|c|c|}
\hline
Field & DOF & $\chi^2$ & Significance\\
\hline
VSA1E    &4778 & 4851 & 0.23 \\ 
VSA1F    &3871 & 3964 & 0.14 \\
VSA1G    &4070 & 3752 & 0.99 \\
VSA2E    &5002 & 5170 & 0.05 \\
VSA2F    &3831 & 3763 & 0.78 \\
VSA2G    &4314 & 4297 & 0.58\\
VSA3E    &4937 & 5011 & 0.22 \\
VSA3F    &4300 & 4325 & 0.39 \\
VSA3G    &4970 & 5060 & 0.18 \\ 
\hline
Day/Night &1051&  986 & 0.92 \\
\hline
\end{tabular}
\end{center}
\end{table}

\section{Power spectrum}

\begin{figure}
\epsfig{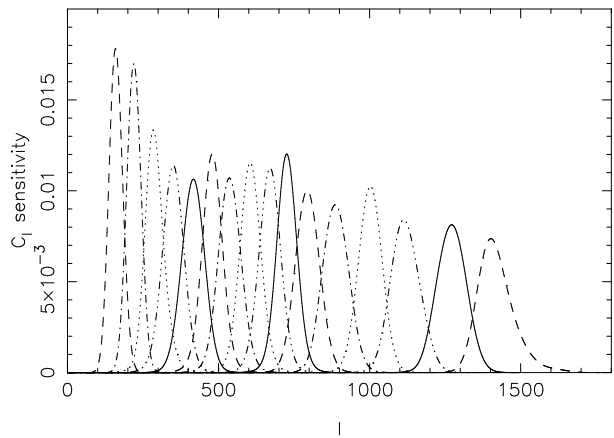}
\caption{Window functions for the combined compact and extended array
data set. The functions are normalised to unit area, and different
bins are plotted with different linestyles. \label{fig:win}}
\end{figure}

\begin{table}
\begin{tabular}{ccccc}
\hline
$B$ & $\ell$-range & $ \ell_{h}$ &  $T_0^2\ell(\ell+1)C_{\ell}/2\pi [\mu {\rm K}^2]$ \\
\hline 
1 & $100 - 190 $ & $160$ &  $3864 _{ -1141} ^{ +1587  }$ &\\
1A & $145 -  220 $ & $190$ & &$4953 _{-1190} ^{+1538} $ \\
2 & $190 - 250 $ & $220$ &  $5893 _{ -1339} ^{ +1637  }$ &\\
2A & $220 -  280 $ & $251$ & &$7031 _{-1339} ^{+1587} $ \\
3 & $250 - 310 $ & $289$ &  $5390 _{ -1091} ^{ +1289  }$ &\\
3A & $280 -  340 $ & $321$ & &$3450 _{-744} ^{+942} $ \\
4 & $310 - 370 $ & $349$ &  $2603 _{ -545} ^{   +595  }$ &\\
4A & $340 -  410 $ & $376$ & &$1954 _{-396} ^{+446} $ \\
5 & $370 - 450 $ & $416$ &  $1749 _{ -347} ^{   +347  }$ &\\
5A & $410 -  475 $ & $431$ & &$1597 _{-347} ^{+446} $ \\
6 & $450 - 500 $ & $479$ &  $1638 _{ -496} ^{   +644  }$ &\\
6A & $475 -  540 $ & $501$ & &$2675 _{-545} ^{+644} $ \\
7 & $500 - 580 $ & $537$ &  $2866 _{ -545} ^{   +595  }$ &\\
7A & $540 -  610 $ & $581$ & &$2210 _{-496} ^{+644} $ \\
8 & $580 - 640 $ & $605$ &  $1460 _{ -545} ^{   +644  }$ &\\
8A & $610 -  670 $ & $639$ & &$1501 _{-595} ^{+595} $ \\
9 & $640 - 700 $ & $670$ &  $2237 _{ -595} ^{   +694  }$ &\\
9A & $670 -  725 $ & $696$ & &$2328 _{-644} ^{+793} $ \\
10 & $700 - 750 $ & $726$ &  $1922 _{ -744} ^{   +793  }$& \\
10A & $725 -  800 $ & $759$ & &$2905 _{-644} ^{+744} $ \\
11 & $750 - 850 $ & $795$ &  $3587 _{ -644} ^{   +644  }$& \\
11A & $800 -  900 $ & $843$ & &$2876 _{-644} ^{+694} $ \\
12 & $850 - 950 $ & $888$ &  $1471 _{ -545} ^{   +644  }$& \\
12A & $900 - 1000$ & $948$ & &$207 _{-207} ^{+694} $ \\ 
13 & $950 - 1050$ & $1002$ & $0 _{     -0} ^{  +1091  }$ &\\
13A & $1000 - 1125$ & $1057$ & &$509 _{-446} ^{+545} $ \\ 
14 & $1050 - 1200$ & $1119$ & $1125 _{ -644} ^{   +694  }$& \\
14A & $1125 - 1275$ & $1199$ & &$590 _{-590} ^{+793} $ \\ 
15 & $1200 - 1350$ & $1271$ & $0 _{     -0} ^{  +1431  }$ &\\
15A & $1275 - 1525$ & $1357$ & &$505 _{-496} ^{+1190} $ \\
16 & $1350 - 1700$ & $1419$ & $1311 _{ -1289} ^{  +1538  }$&\\

\hline
\end{tabular}
\caption {The bandpowers for main and offset binnings of the complete
VSA data set, combining both compact and extended array data.  The
$\ell$-range gives the nominal bin limits, while $\ell_h$ is the
median value of the relevant window function. The error bars enclose
68 percent of the likelihood, even when the lower limit is
zero. \label{bandpowers}}
\end{table}

\begin{table}
\begin{tabular}{ccccccc}
$B$ & ${\cal C}_{B,B-2}$ & ${\cal C}_{B,B-1}$ & ${\cal C}_{B,B}$ &
${\cal C}_{B,B+1}$ & ${\cal C}_{B,B+2}$ & $\mbox{Cov}_{B,B} $\\
\hline
$ 1 $ &  &    &   $ 1.0 $  &   $ -0.09 $  &   $ 0.012 $ & $ 17.039 $ \\
$ 2 $ &  &   $ -0.09 $  &   $ 1.0 $  &   $ -0.206 $  &   $ 0.021 $ & $ 22.295 $ \\
$ 3 $ & $ 0.012 $  &   $ -0.206 $  &   $ 1.0 $  &   $ -0.114 $  &   $ 0.021 $ & $ 14.886 $ \\
$ 4 $ & $ 0.021 $  &   $ -0.114 $  &   $ 1.0 $  &   $ -0.151 $  &   $ 0.031 $ & $ 3.442 $ \\
$ 5 $ & $ 0.021 $  &   $ -0.151 $  &   $ 1.0 $  &   $ -0.227 $  &   $ 0.044 $ & $ 1.316 $ \\
$ 6 $ & $ 0.031 $  &   $ -0.227 $  &   $ 1.0 $  &   $ -0.254 $  &   $ 0.047 $ & $ 3.584 $ \\
$ 7 $ & $ 0.044 $  &   $ -0.254 $  &   $ 1.0 $  &   $ -0.268 $  &   $ 0.037 $ & $ 3.912 $ \\
$ 8 $ & $ 0.047 $  &   $ -0.268 $  &   $ 1.0 $  &   $ -0.211 $  &   $ 0.09 $ & $ 4.011 $ \\
$ 9 $ & $ 0.037 $  &   $ -0.211 $  &   $ 1.0 $  &   $ -0.406 $  &   $ 0.092 $ & $ 5.135 $ \\
$ 10 $ & $ 0.09 $  &   $ -0.406 $  &   $ 1.0 $  &   $ -0.286 $  &   $ 0.055 $ & $ 7.863 $ \\
$ 11 $ & $ 0.092 $  &   $ -0.286 $  &   $ 1.0 $  &   $ -0.187 $  &   $ 0.049 $ & $ 4.883 $ \\
$ 12 $ & $ 0.055 $  &   $ -0.187 $  &   $ 1.0 $  &   $ -0.251 $  &   $ 0.079 $ & $ 3.905 $ \\
$ 13 $ & $ 0.049 $  &   $ -0.251 $  &   $ 1.0 $  &   $ -0.336 $  &   $ 0.052 $ & $ 2.353 $ \\
$ 14 $ & $ 0.079 $  &   $ -0.336 $  &   $ 1.0 $  &   $ -0.162 $  &   $ 0.03 $ & $ 5.475 $ \\
$ 15 $ & $ 0.052 $  &   $ -0.162 $  &   $ 1.0 $  &   $ -0.196 $  &   & $ 6.714 $ \\
$ 16 $ & $ 0.03 $  &   $ -0.196 $  &   $ 1.0 $  &    &   & $ 27.092 $ \\
\end{tabular}
\caption{The correlation matrix ${\cal C}_{i,j}$ for the combined VSA
data (main binning only). The values of the matrix for which ${\cal
C}_{i,j}$ is not reported can be assumed 
to be zero. The final column gives the diagonal elements of the covariance
matrix in units of $10^5 \times \mu \rm{K}^4 $\label{corrmat}}
\end{table}

The fully reduced and source subtracted visibility data for the combined
compact and extended VSA data sets have been analysed using the {\sc Madcow}
software package \citep{klaus}.  
The typical anti-correlation between adjacent bins is
$\sim20$~percent, with one as high as $40$~percent. 
We use variable
sized bins to reflect the differences in density of coverage of the aperture
plane, but also repeat the calculation with bin centres shifted by half a bin
width in order to sample effectively the features of the power spectrum.  We
follow the procedure outlined in Paper\,III to determine the window function
$W(\ell)$ which determines how each bin samples the underlying power
spectrum. The window functions for our analysis are shown in
Figure~\ref{fig:win}. The resulting power spectrum is shown in
Figure~\ref{CMB_doublebin}, numerical values for both binnings are given in
Table~\ref{bandpowers} and the correlation matrix for the main binning in
Table~\ref{corrmat}. The bin centres given in each case are the medians of the
respective window functions rather than the nominal bin centres.

There is a systematic uncertainty of 7 percent in the scaling of the
power spectrum due to the uncertainty in the effective temperature of
the primary calibration source, Jupiter. The error in flux density to
temperature conversion (equivalent to the beam uncertainty in
scanned-beam experiments) is negligible.

\section{Discussion \& conclusions}

\begin{figure*}
\centering
\begin{minipage}{1\textwidth}
\centering
\epsfig{file=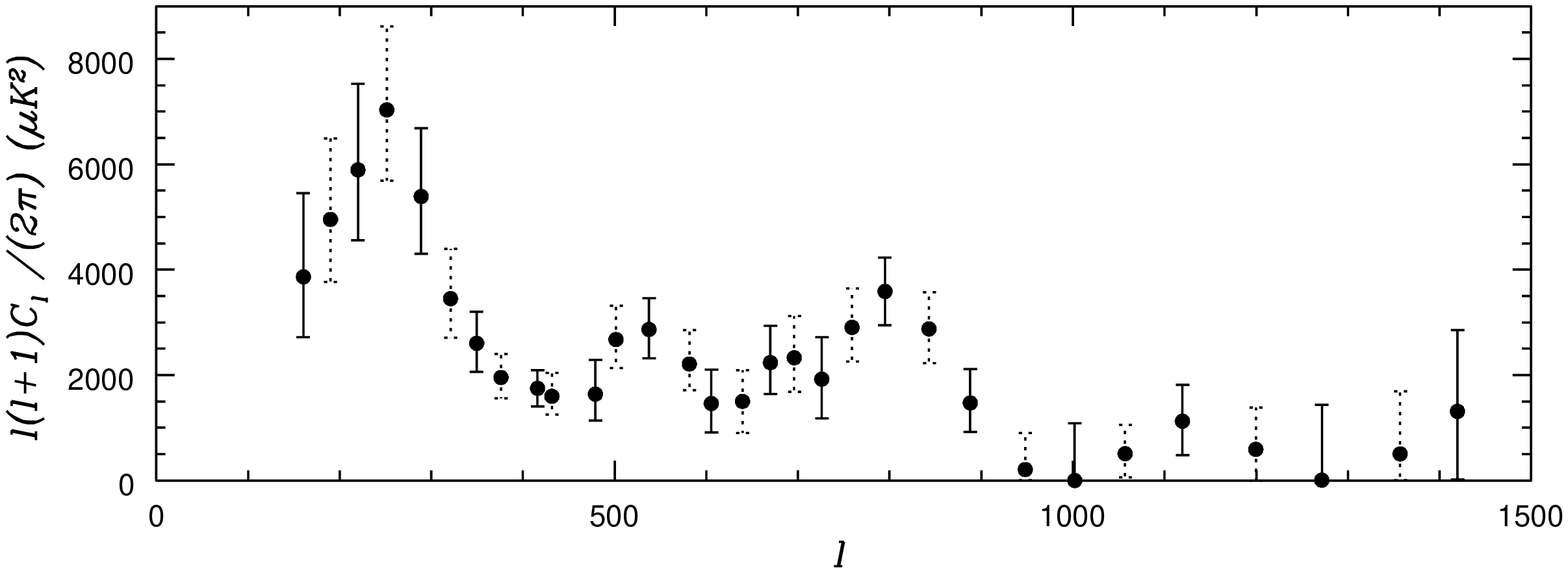,height=170mm,angle=270,clip=}
\caption{Combined CMB power spectrum from the three mosaiced
     VSA fields.  The error-bars represent $1\sigma$ limits; the two sets of
      data points correspond to alternative interleaved binnings of the data. \label{CMB_doublebin}}
\end{minipage}\\[20pt]
\begin{minipage}{1\textwidth}
\centering
\epsfig{file=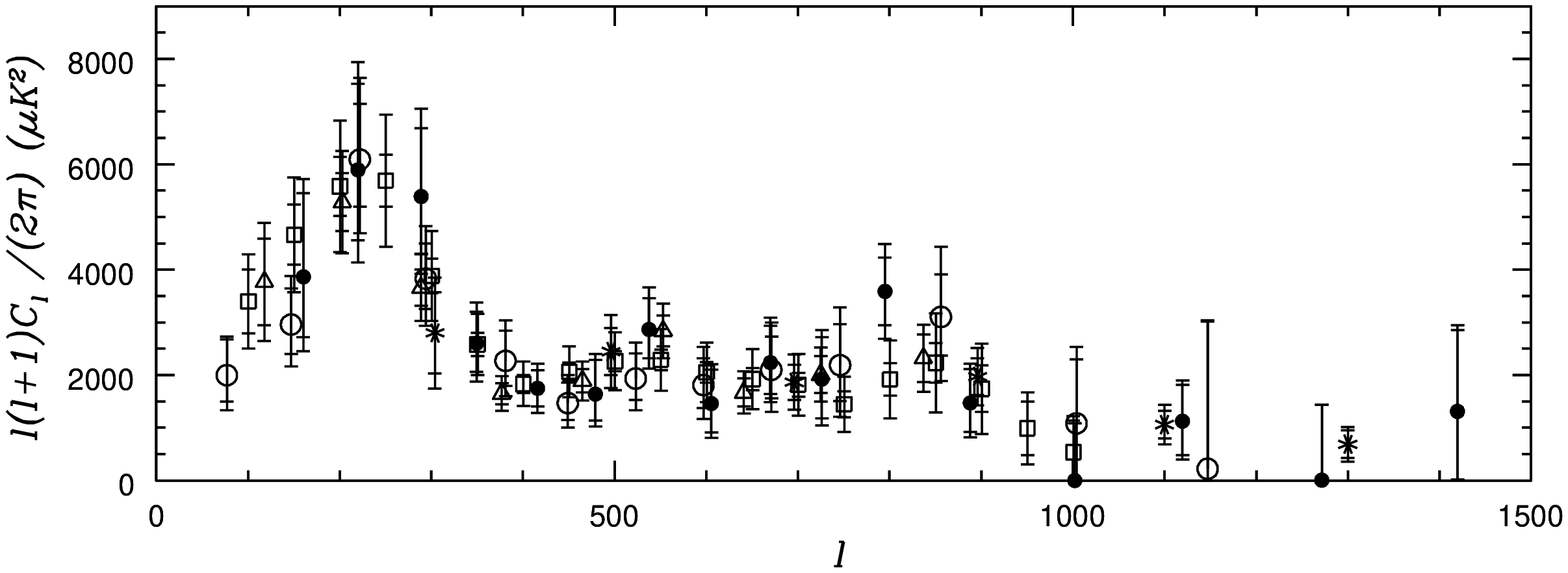,height=170mm,angle=270,clip=}
\caption{A comparison of the VSA data (filled circles) with results from 
  the BOOMERANG (open squares), MAXIMA (open circles), DASI (open
  triangles) and CBI (stars) experiments. 
Two sets of error bars are plotted for each
  data set; the smaller (inner) of the two indicate only random errors, whilst
  the larger bars indicate the amount by which the inner points could move
  due to absolute calibration and beam uncertainty. In each case the
  error bars indicate $1\sigma$ limits.\label{CMB_comparison}}
\end{minipage}
\end{figure*}

We have measured the power spectrum of the CMB over the multipole range $\ell
= 160$ -- $1400$ using two configurations of the VSA. The combined power
spectrum clearly shows three peaks, a sharp fall-off in
power above $\ell~\sim~800$, and marginal evidence for a fourth
peak. Figure~\ref{CMB_comparison} shows the VSA data plotted alongside
that from 
other recent CMB experiments: there is excellent agreement between all the
experiments.

To quantify the significance of these peaks in the power spectrum, we used a
model of five Gaussians parametrised by their amplitude, position and width,
as described in \citet{Odman2002}.
We analysed the VSA data alone, plus two data sets consisting of a compilation
of recent data, and used an MCMC routine to constrain the parameters
simultaneously. In Table~\ref{constraints}, we list the constraints on the
amplitudes and positions of the first four peaks after marginalising
over their widths (there are no interesting
constraints on the fifth peak). Including the new VSA measurement improves the
constraints on the amplitude of the fourth peak, as well as confirming the
detection of the first three peaks.

Fits of adiabatic inflationary model power spectrum, with associated
estimates of cosmological parameters, are presented in a companion paper
\citep{VSApaperVI}.

In our current observing configuration we are increasing the size of the
mosaiced regions and also observing new regions; this will increase both the
sensitivity and $\ell$-resolution of the power spectrum. We also plan
to increase the $\ell$-range of 
the VSA using larger antennas and longer baselines to study further peaks in
the primordial spectrum and give information on the region where CBI observes
excess power~\citep{mason-02}.

Band powers, correlation matrices and window functions can be downloaded from\\
\noindent \verb+http://mrao.cam.ac.uk/telescopes/vsa/results.html+

\begin{table}
\begin{center}
\caption{Constraints on parameters of a multiple Gaussian fit to the
power spectrum data.
The inclusion of the new VSA data places constraints on the power beyond
the first three peaks. Dataset I consists of the QMASK
compilation~\citep{Xu2002}, MSAM~\citep{Wilson2000}, 
Python V~\citep{Coble2001}, Boomerang 98~\citep{netterfield-01},
Maxima~\citep{lee-01}, DASI~\citep{halverson-02}, CBI~\citep{pearson-02},
Archeops~\citep{Benoit2002} and the VSA data from the
compact array configuration. Dataset II is identical, except for the inclusion
of the new VSA data.
Note that, while the lack of a lower bound on a $\Delta$T implies a
significant contribution to the power from adjacent Gaussians, narrow
constraints on peak locations imply strong evidence for spectral
features rather than a flat spectrum of excess power.\label{constraints}}
\begin{tabular}{|l|c|c|c|}\hline
 & VSA only & dataset I & dataset II \\ \hline
$\Delta T_1$ & $71.4_{-5.5}^{+11.3}$  & $70.2_{-2.0}^{+2.1}$  &
$71.7_{-2.1}^{+2.0}$\\
$\Delta T_2$ & $45.5_{-5.6}^{+6.4}$  &
$38.5_{-7.8}^{+8.1}$  & $42.3_{-10.4}^{+4.7}$\\
$\Delta T_3$ &
$47.0_{\mbox{\tiny unbounded}}^{+8.9}$  & $42.5_{-2.2}^{+1.5}$ &
$43.3_{-2.3}^{+1.7}$ \\
$\Delta T_4$ & $25.9_{\mbox{\tiny
unbounded}}^{+14.1}$  & $18.0_{\mbox{\tiny unbounded}}^{+19.1}$ &
$19.6_{-11}^{+11.7}$ \\
$\ell_1$ & $224_{-28}^{+11}$  &
$214_{-5}^{+4}$  & $216_{-6}^{+3}$\\
$\ell_2$ & $509_{-84}^{+81}$  &
$511_{-16}^{+12}$  & $510_{-14}^{+11}$\\
$\ell_3$ & $771_{-70}^{+55}$  &
$786_{-48}^{+29}$  & $782_{-52}^{+23}$\\
$\ell_4$ & $1171_{-205}^{+67}$  &
$1260_{-321}^{+37}$  & $1157_{-118}^{+167}$\\
\hline
\end{tabular}
\end{center}
\end{table}

\section*{ACKNOWLEDGEMENTS}

We thank the staff of MRAO, JBO and the Teide Observatory for invaluable
assistance in the commissioning and operation of the VSA. The VSA is supported
by PPARC and the IAC. Partial financial support was provided by the Spanish
Ministry of Science and Technology.  CD, RS and KL acknowledge support by PPARC
studentships. KC acknowledges a Marie Curie Fellowship.  
YH is supported by the Space Research Institute of KACST.
AS acknowledges the
support of St. Johns College, Cambridge. We thank Professor Jasper Wall for
assistance and advice throughout the project.

\label{lastpage}
\bibliography{cmb_refs} \bibliographystyle{mn2e} \bsp \end{document}